\documentclass[twocolumn,prb,showpacs,preprintnumbers,amsmath,amssymb]{revtex4}
\usepackage{graphicx}    
\usepackage{dcolumn}     
\usepackage{bm}          
\newcommand{\be}{\begin{equation}}
\newcommand{\ee}{\end{equation}}
\newcommand{\bes}{\begin{equation*}}
\newcommand{\ees}{\end{equation*}}
\newcommand{\bea}{\begin{eqnarray}}
\newcommand{\eea}{\end{eqnarray}}
\newcommand{\beas}{\begin{eqnarray*}}
\newcommand{\eeas}{\end{eqnarray*}}
\newcommand{\ba}{\begin{array}}
\newcommand{\ea}{\end{array}}

\begin{document}

\title{Multiple magnetic barriers in graphene}

\author{Luca Dell'Anna$^1$ and Alessandro De Martino$^{2,3}$}

\affiliation{
$^1$ International School for Advanced Studies, SISSA-ISAS, 
I-34014 Trieste, Italy\\
$^2$ Institut f\"ur Theoretische Physik, Universit\"at zu K\"oln, 
D-50937 K\"oln, Germany\\
$^3$ Institut f\"ur Theoretische Physik, Heinrich-Heine-Universit\"at,
D-40225  D\"usseldorf, Germany}

\date{\today}

\begin{abstract}

We study the behavior of charge carriers in graphene in inhomogeneous
perpendicular magnetic fields. We consider two types of one-dimensional
magnetic profiles, uniform in one direction:
a sequence of $N$ magnetic barriers, and a sequence of 
alternating magnetic barriers and wells. In both cases, we compute the 
transmission coefficient of the magnetic structure by means of the transfer 
matrix formalism, and the associated conductance.
In the first case the structure becomes increasingly transparent
upon increasing $N$ at fixed total magnetic flux.
In the second case we find strong wave-vector filtering and resonant effects.
We also calculate the band structure of a periodic magnetic 
superlattice, and find a wave-vector-dependent gap around zero-energy.
\end{abstract}

\pacs{73.21.-b, 73.63.-b, 75.70.Ak}

\maketitle

\section{Introduction}
\label{intro}

The electronic properties of graphene\cite{reviews} 
in the presence of inhomogeneous perpendicular magnetic fields
have very recently attracted considerable theoretical attention 
\cite{ale,tarun,cserti,sim,masir,tahir,zhai,heinzel,kormanios,ghosh}. 
In graphene the charge carriers close to the Fermi points $K$ 
and $K'$ form a relativistic gas of chiral massless (Dirac-Weyl)
quasiparticles with a characteristic conical spectrum.
This has far-reaching consequences. For example,
quasiparticles in graphene are able to tunnel through 
high and wide electrostatic potential barriers,
a phenomenon often referred to as Klein tunneling, 
and related to their chiral nature\cite{kleintunn}. 
Moreover, 
in a uniform magnetic field graphene exhibits
an unconventional half-integer quantum Hall effect
\cite{iqheexp}, which 
can be understood in terms of the existence, among 
the relativistic Landau levels formed by the quasiparticles, 
of a zero-energy one\cite{iqheth}.

From a theoretical perspective, it is then interesting
to explore how the Dirac-Weyl (DW) nature of the charge carriers 
affects their behavior in non-uniform magnetic fields. 
Such investigation has been started in Ref. \cite{ale},
and here we generalize and expand on it 
by studying several more complex geometries.

Experimentally, inhomogeneous magnetic profiles 
on submicron scales in ordinary 2DEGs 
in semiconductor heterostructures have been 
produced in several ways, and magnetic barriers 
with heights up to 1 T have been obtained. 
One approach exploits the fringe field produced 
by ferromagnetic stripes fabricated on top of 
the structure \cite{heinzelex}.
Another possibility consists in applying a uniform magnetic field 
to a 2DEG with a step \cite{leadbeater}.
In yet another approach, a film of superconducting 
material with the desired pattern is deposited 
on top of the structure, and a uniform magnetic field is applied.\cite{bending}
In this way, magnetic structures with different geometries 
have been  experimentally realized, and their 
mesoscopic transport properties have been studied,
e.g. transport through single magnetic barriers \cite{expbar} and
superlattices \cite{carmona}, magnetic edge states 
close to a magnetic step \cite{henini}, and 
magnetically confined quantum dots or antidots \cite{antidots}.
Correspondingly, there exists an extensive theoretical 
literature, pioneered by the works of F.M. Peeters 
and collaborators \cite{peetersstructures}, which elucidates
the basic mechanisms underlying the behaviors observed in 
experiments. 

In principle, the same concepts and technologies 
can be used to create similar magnetic structures in graphene, 
once the graphene sheet is covered by an insulating layer,
which has recently been demonstrated feasible \cite{huard,williams}.
Although at the time of writing there is yet no published 
experimental work demonstrating magnetic barriers in graphene, 
this should be within reach of present-day technology, 
which provides motivation for the present work.

In a previous paper \cite{ale} we showed that, in contrast 
to electrostatic barriers, a single magnetic barrier in graphene 
totally reflects an incoming electron, provided the electron 
energy does not exceed a threshold value related to the total 
magnetic flux through the barrier. Above this threshold,  the 
transmission coefficient strongly depends on the incidence 
angle \cite{ale, masir}. These observations were used to argue 
that charge carriers in graphene can be confined by means of 
magnetic barriers, which may thus provide efficient tools to 
control the transport properties in future graphene-based 
nanodevices.

Here we focus on more complex multiple barrier 
configurations and magnetic superlattices. 
We consider two types of one-dimensional
profiles. In the first case the magnetic field in the 
barrier regions is always assumed to point upwards,  
while in the second it points alternatingly upwards 
and downwards. We shall see that there are sharp 
differences in the transport properties of the two cases.

The outline of the paper is the following. 
In Sec. \ref{sec2} we introduce the Dirac-Weyl Hamiltonian 
for graphene, the two types of magnetic profiles we consider 
in the rest of the paper, and the transfer matrix 
formalism for Dirac-Weyl particles.
In Sec. \ref{sec3} and Sec. \ref{sec4} we
compute and discuss the transmission coefficient separately  
for the two cases.
In Sec. \ref{sec5} we consider a periodic magnetic superlattice, 
and determine its band structure.
Finally, in Sec. \ref{sec6} we summarize our results 
and draw our conclusions.

\section{Hamiltonian and transfer matrix}
\label{sec2}

Electrons in clean graphene close to the two Fermi points $K$ and $K'$
are described by two decoupled copies of the Dirac-Weyl (DW) equation.  
We shall focus here on a single valley and neglect the 
electron spin.\cite{footnote1} Including the perpendicular 
magnetic field via minimal coupling, the DW equation reads 
\begin{equation}
v_F {\bm \sigma} \cdot \left(-i{\bf \hbar \nabla} + \frac{e}{c} 
{\bf A}\right) \Psi = E \Psi ,
\label{dirac2d}
\end{equation}
where ${\bm \sigma}= (\sigma_x,\sigma_y)$ are Pauli matrices acting
in sublattice space, and $v_F=8\times10^5\,$m/s is the 
Fermi velocity in graphene. 
In the Landau gauge, 
${\bm A}=(0, A(x))$, with $B_z = \partial_x A$, the $y$-component of the
momentum is a constant of motion, and the spinor
wavefunction can be written as $\Psi(x,y) = \psi(x)e^{ik_yy}$, 
whereby Eq. (\ref{dirac2d})  is reduced to a one-dimensional problem:
\begin{equation}
\left( 
\begin{array}{cc}
- E & -i\partial_x -i(k_y + A(x)) \\
-i\partial_x +i(k_y+A(x)) & - E
\end{array}
\right) \psi = 0.
\label{dirac1d}
\end{equation}
Eq. (\ref{dirac1d}) is written in dimensionless units: with $B$ 
the typical magnitude of the magnetic field, and 
$\ell_B=\sqrt{\hbar c/e B}$ the associated
magnetic length, we express the vector
potential $A(x)$ in units of $B\ell_B$, the energy $E$ 
in units of $\hbar v_F/\ell_B$, and $x$ and $k_y$ respectively 
in units of $\ell_B$ and  $\ell^{-1}_B$. 
The values of local magnetic fields 
in the barrier structures produced by
ferromagnetic stripes range up to 1 T, 
with typical values of the order of tenth of Tesla.
For $B\approx 0.1\,$T, we find $\ell_B \approx 80\,$nm and  
$\hbar v_F/\ell_B \approx 7\,$meV, which set the 
typical length and energy scales.

\begin{figure}[h!]
\includegraphics[width=0.32\textwidth]{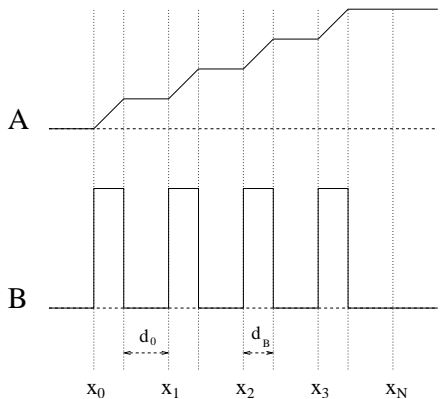}
\caption{Magnetic profile Eq. (\ref{gauge1}): 
$N$ magnetic barriers of width $d_B$ separated by 
nonmagnetic regions of width $d_0$.}
\label{profile1}
\end{figure}

We shall consider two types of magnetic field profiles. 
In the first case, illustrated in Fig. \ref{profile1},
the profile consists of a sequence of $N$ magnetic barriers 
of equal height $B$ (assumed positive for definiteness) 
and width $d_B$, separated by nonmagnetic 
regions of width $d_0$. The vector potential is then chosen as 
\begin{equation}
A(x) = \left\{
\begin{array}{ll}
0,&  \;\; x\in [-\infty,0],\\
n d_B + (x-x_{n}), &  \;\; x\in [x_{n}, x_{n}+ d_B] ,\\
(n+1) d_B, &   \;\; x\in [x_{n}+ d_B, x_{n+1}] ,\\
N  d_B, &  \;\; x\in [x_{N},\infty],
\end{array}
\right.
\label{gauge1}
\end{equation}
where $n=0,\dots, N-1$ and $x_{n}=n (d_0+d_B)$. The 
quantity $Nd_B$ is the total magnetic flux through 
the structure per unit length in the $y$-direction.\cite{footnote2}
We shall refer to this profile as the 
multiple barriers case and discuss it in Sec. \ref{sec3}.

\begin{figure}[h!]
\includegraphics[width=0.32\textwidth]{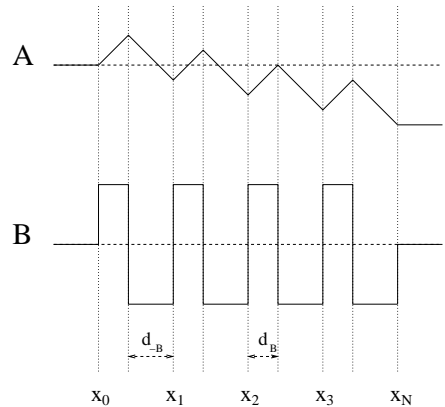}
\caption{Magnetic profile Eq. (\ref{gauge2}): $N$ magnetic barriers of width 
$d_B$ separated by magnetic wells of width $d_{-B}$.}
\label{profile2}
\end{figure}

In the second case, illustrated in Fig. \ref{profile2}, 
each magnetic barrier is followed by
a region of width $d_{-B}$ of opposite magnetic field.
The vector potential is accordingly
chosen as 
\begin{equation}
A(x) = \left\{
\begin{array}{ll}
0, &  \;\; x\in [-\infty,0],\\
n D +(x-x_{n}), &  \;\; x\in [x_{n}, x_{n}+ d_B], \\
n D+(2d_{B}+x_{n}-x), &   \;\; x\in [x_{n}+d_B, x_{n+1}], \\
N D, &   \;\; x\in [x_{N},\infty],
\end{array}
\right.
\label{gauge2}
\end{equation}
where $n=0,\dots,N-1$, $x_{n}=n (d_B+d_{-B})$, and $D=d_B-d_{-B} $. 
We shall refer to this profile as the alternating 
barrier-well case, and discuss it in 
Sec. \ref{sec4}. The parameter $D$ has the meaning 
of net magnetic flux through a cell formed by a barrier 
and a well. For $D=0$ this profile can be extended to a 
periodic magnetic superlattice, a case considered in 
Sec. \ref{sec5}.

With our gauge choice, the value of the vector potential 
on the right of the structure is equal to
the total magnetic flux $\Phi$ through it 
\be 
\Phi \equiv A(x>x_N)=\left\{ \begin{array}{ll}
Nd_B, & \quad {\text{case 1}},\\
N(d_B-d_{-B}), & \quad {\text{case 2}},
\end{array} \right. 
\label{flux}
\ee
which is an important control parameter for the transport 
properties. 

In both cases, the solutions to Eq. (\ref{dirac1d}) can be obtained
by first writing the general solution in each 
region of constant $B_z$ as linear combination 
(with complex coefficients) of the
two independent elementary solutions, 
and then imposing the continuity of the wavefunction 
at the interfaces between regions of different $B_z$, 
to fix the complex coefficients. 
This procedure is most conveniently performed in the 
transfer matrix formalism. Here we directly use this approach 
and refer the reader to Refs.\cite{mckellar}, \cite{barbier} 
for a detailed discussion.

 The transfer matrix 
\be
\hat T =\left(
\ba{cc}
T_{11}&T_{12}\\
T_{21}&T_{22}
\ea
\right)
\ee
relates the wave function on the left side of the magnetic structure
($x<x_0=0$)
\be
\psi(x)=\left(
\begin{array}{l}
1 \\
\frac{k^i_x+ik_y}{E}
\end{array}
\right) e^{i k^i_x x} +
r\left(
\begin{array}{l}
1 \\
\frac{-k^i_x+ik_y}{E}
\end{array}
\right) e^{-i k^i_x x},
\ee
where $k^i_x = \sqrt{E^2-k_y^2}$,
 to the wave function on the right side ($x>x_N$)
\be
\psi(x) = t \sqrt{\frac{k^i_x}{k^f_x}}\left(
\begin{array}{c}
1 \\ \frac{k^f_x+i(k_y+\Phi)}{E} \end{array}
\right) e^{ik^f_x x} ,
\ee
where $k_x^f=\sqrt{E^2-(k_y+\Phi)^2}$.
The coefficients $r$ and $t$ are resp. the reflection 
and transmission amplitudes, and
we used that with our gauge choices (\ref{gauge1}) and (\ref{gauge2}),
the vector potential vanishes on the left of the magnetic structure, 
and on the  right side is equal to $\Phi$.
As usual, the factor $\sqrt{k^i_x/k^f_x}$ ensures 
proper normalization of the probability current.
The relation which expresses the continuity of the 
wave function is then 
given by\cite{transfer} 
\begin{equation}
\left(
\begin{array}{l}
1\\r \end{array}
\right)
=\hat{T}
\left(
\begin{array}{l}
\sqrt{k^i_x/k^f_x}\,t\\0 \end{array}
\right).
\label{rt}
\end{equation}
Solving Eq. (\ref{rt}) for $t$, we get the transmission probability 
${\cal T}$ as
\be
{\cal T}(E,k_y)=|t|^2=\frac{k_x^f}{k_x^i}\frac{1}{|T_{11}|^2}.
\label{transm}
\ee
Once ${\cal T}(E,k_y)$ is known, it is straightforward 
to compute the zero-temperature
conductance by integrating ${\cal T}$ over one half of the 
(Fermi) energy surface\cite{peetersfilter,masir}:
\begin{equation}
G(E) 
= G_0 \int_{-\frac{\pi}{2}}^{\frac{\pi}{2}} d\phi \, \cos \phi 
\, {\cal T}(E,E\sin \phi) ,
\label{cond}
\end{equation}
where $\phi$ is the incidence angle (we measure angles with respect
to the $x$-direction), defined by $k_y=E\sin\phi$, and 
$G_0=2e^2EL_y/\pi h$. $L_y$ is the length of the graphene 
sample in the $y$ direction, and $G_0$ includes a 
factor $4$ coming from the spin and valley degeneracy. 

Before to proceed with the calculations, 
we can derive a simple and general condition for 
a non-vanishing transmission. 
For this purpose it is convenient to parametrize the momenta 
in the leftmost and rightmost regions resp. in terms of 
incidence and emergence angles, $\phi$ and $\phi_f$:
\begin{eqnarray}
k^i_x&=&E \cos \phi, \quad \quad k_y= E \sin \phi,\\
k^f_x&=&E \cos \phi_f, \quad \quad k_y= E \sin \phi_f-\Phi.
\end{eqnarray}
The emergence angle is then fixed by the conservation of  $k_y$:
\be
\sin \phi= \sin \phi_f - \frac{\Phi}{E} .
\label{mastereq}
\ee
Equation (\ref{mastereq}) implies that transmission through 
the structure is only possible if $\phi_i$ satisfies the condition
\be
\left| \sin\phi + \frac{\Phi}{E} \right| \le 1.
\label{threshold}
\ee
This condition, already discussed in Ref. \cite{ale} for the case 
of a single barrier, is in fact completely general and
independent of the detailed form of the magnetic field profile.
It only requires that the magnetic field vanishes outside 
a finite region of space. For $|\Phi/E|>2$, it implies that the 
magnetic structure completely 
reflects both quasiparticles and quasiholes. 
As a consequence of this angular threshold, the conductance 
has an upper bound given by
\be
\label{Gs}
G_s(E)\equiv G_0 \left(2-\left|{\frac{\Phi}{E}}
\right|\right) \;\theta\left(2|E|-|\Phi|\right),
\ee
with the Heaviside step function $\theta$. If the 
vector potential profile is monotonous,
$G_s$ also coincides with the classical conductance, 
obtained by setting ${\cal T}=\theta(1-|\sin \phi +\Phi/E|)$. 
If, however, $A(x)$ is not monotonous, the classical conductance 
is obtained by replacing $|\Phi|$ in Eq. (\ref{Gs})  
with the maximal value of $|A|$ in the
structure, since a classical particle is totally 
reflected as soon as $|A|_{max}$ (rather than the 
total flux $|\Phi|$) exceeds twice the energy.

Before moving to the next section, we notice, as an aside remark, 
that Eq. (\ref{dirac1d}) can easily be solved in closed form for $E=0$. 
The zero-energy spinors are then given by 
\bea
\psi_+ &\propto&  \left(
\begin{array}{c}
1\\
0
\end{array} \right)
e^{k_y x +\int^x A(x')dx'}, \label{e0u}\\
\psi_- &\propto&  
\left(
\begin{array}{c}
0\\
1
\end{array}\right)
e^{-k_y x -\int^x A(x')dx'}. \label{e0v} 
\eea
These wave functions are admissible if and
only if they are normalizable,
which depends on the sign of $k_y$ and the behavior of the 
magnetic field at $x\rightarrow \pm \infty$. 
In fact, for any $A(x)$, at most one among
 $\psi_+$ and $\psi_-$ is admissible. 
If the magnetic field vanishes outside a 
finite region of space, as in our case, one can 
always choose a gauge in which $A(x)=0$ on the left
of the magnetic region and $A(x)=\Phi $ on its right.
It is then straightforward to check that for
$0<k_y<-\Phi$, the only normalizable
solution is (\ref{e0u}), whereas for
$-\Phi<k_y<0$, the normalizable solution is (\ref{e0v}).
In particular, we find that, when the net magnetic flux through the 
structure vanishes, there exist no zero-energy states. 
This is nicely confirmed by the calculation of the 
spectrum of the periodic magnetic superlattice in Sec. \ref{sec5}. 
The zero-energy state is a bound state localized in the structure.
Additional bound states of higher energy may also
occur\cite{masir}, but we do not further investigate this problem here.

\section{Multiple barriers}
\label{sec3}

In this section we focus on the magnetic profile in Eq. (\ref{gauge1}). 
In order to compute the transfer matrix $\hat T$, we need 
the two elementary solutions of the DW equation (\ref{dirac1d}) 
for $B_z=0$ and the two for $B_z=1$. 

\begin{figure}[h]
\includegraphics[width=0.44\textwidth]{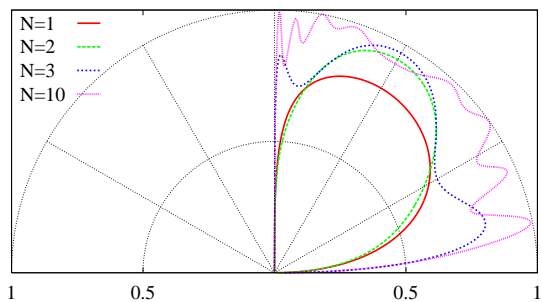}
\caption{(Color online) Angular dependence of the transmission probability 
at $E=1$, through $N=1,2,3,10$ barriers of 
width $d_B=1/N$ (keeping in this way constant the flux $\Phi=Nd_B=1$), 
and spaced by $d_0=10$.}
\label{tran1_N}
\end{figure}

\begin{figure}[h]
\includegraphics[width=0.44\textwidth]{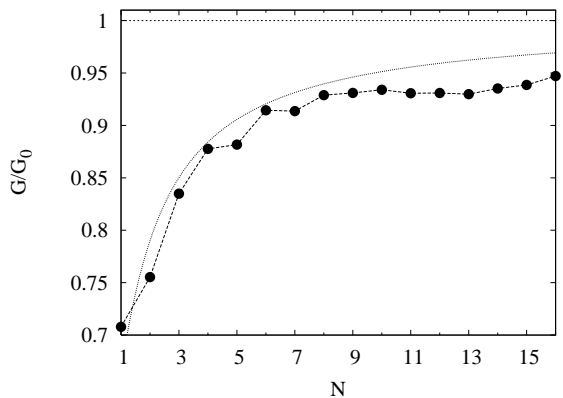}
\caption{The conductance $G/G_0$ at $E=1$ 
for several values of $N$ with $d_B=1/N$, such that $\Phi=1$, and $d_0=10$ (black dots linked by long-dashed line, which is a guide to the eye only). 
The upper bound (dashed line) corresponds to the classical value 
$G_s/G_0=1$  (see Eq. (\ref{Gs})) while the dotted line is the curve given 
by Eq. (\ref{G_approx}) as a function of $N$.}
\label{cond1_N}
\end{figure}

We can then construct the $2\times 2$ matrices ${\cal W}_0(x)$ 
and ${\cal W}_B(x)$, whose columns are given by the 
 spinor solutions. In the nonmagnetic regions we have
\begin{equation}
\label{W_0}
{\cal W}_0(x) =
\left(
\begin{array}{ll}
e^{i k_x x} & e^{-i k_x x}  \\
\frac{k_x+i(k_y+A)}{E}e^{ik_x x} & \frac{-k_x+i(k_y+A)}{E}e^{-i k_x x}
\end{array}
\right),
\end{equation}
where  $k_x(x)=\sqrt{E^2-(k_y+A(x))^2}$.
In the regions with $B_z=1$ we have
\begin{equation}
{\cal W}_B(x) =
\left(
\begin{array}{ll}
D_p(q) & D_p(-q)\\
\frac{i\sqrt{2}}{E}D_{p+1}(q) &\frac{-i \sqrt{2}}{E}D_{p+1}(-q)
\end{array}
\right) ,
\end{equation}
where $q=\sqrt{2}(A(x)+k_y)$, $p=E^2/2-1$, and $D_p(q)$ is 
the parabolic cylinder function\cite{gradshtein}. 
These matrices play the role of partial transfer matrices, and allow us to  
express the condition of continuity of the wave function at each 
interface between the nonmagnetic and the magnetic regions. After 
straightforward algebra, we get 
\begin{equation}
\hat{T} = \hat T_0 \, \hat T_1 \, \cdots \, \hat T_{N-1}, 
\label{t1}
\end{equation}
where
\begin{equation}
\hat T_n = {\cal W}_0^{-1}(x_{n}){\cal W}_B(x_{n})
{\cal W}_B^{-1}(x_{n}+d_B){\cal W}_0(x_n+d_B),
\label{tncase1}
\end{equation}
is the transfer matrix\cite{transfer} 
across the $(n+1)^{\text{\tiny th}}$ barrier, 
and we remind that $x_n=n(d_0+d_B)$.

\begin{figure}[h]
\includegraphics[width=0.44\textwidth]{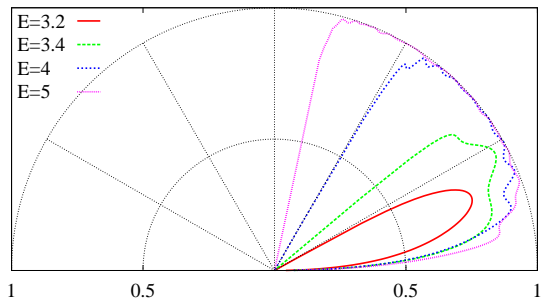}
\caption{(Color online) Angular dependence of the transmission probability, 
for different values of $E$, fixing $d_B=1$, $d_0=10$, and $N=6$.}
\label{tran1_E}
\end{figure}

\begin{figure}[h]
\includegraphics[width=0.44\textwidth]{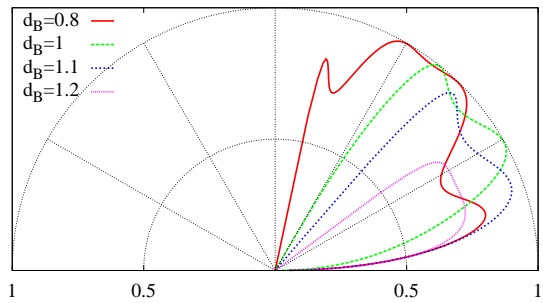}
\caption{(Color online) Angular dependence of the transmission probability, 
for different values of $d_B$, fixing $E=2$, $d_0=10$, and $N=3$.}
\label{tran1_db}
\end{figure}

From Eqs. (\ref{tncase1}), (\ref{t1}), and  (\ref{transm}) \
we numerically evaluated the transmission probability ${\cal T}$ 
for various sets of parameters. The results are illustrated in 
Figs. \ref{tran1_N}, \ref{tran1_E} and \ref{tran1_db}.
Fig. \ref{tran1_N} shows the angular dependence of the 
transmission coefficient at fixed energy for several values 
of $N$, but keeping constant the magnetic flux $\Phi$ 
through the structure. In agreement with the discussion 
in the previous section and Eq. (\ref{threshold}), we observe 
that the range of angles where ${\cal T}\neq 0$ remains the same, 
$\phi \le 0$, upon increasing the number of barriers. 
At the same time, however,  
the transmission itself is modified, and oscillations appear, whose 
number increases with $N$ and for larger separations 
between the barriers.

\begin{figure}[h]
\includegraphics[width=0.44\textwidth]{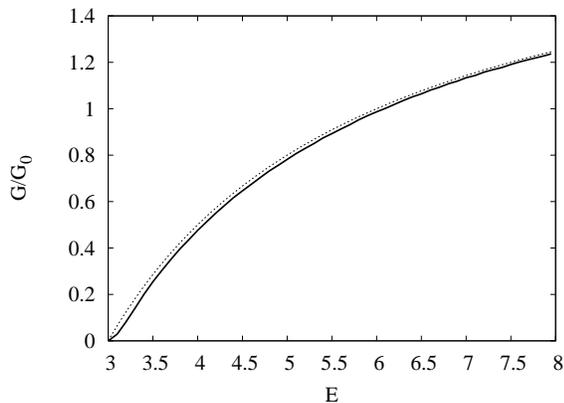}
\caption{The conductance as a function of the energy for $N=6$  
barriers with $d_B=1$ and $d_0=10$ (solid line). The dashed line is the curve 
given by the classical limit Eq. (\ref{Gs}).}
\label{cond1_E}
\end{figure}

\begin{figure}[h]
\includegraphics[width=0.44\textwidth]{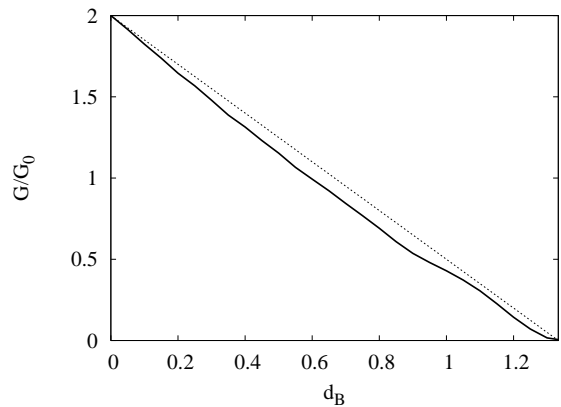}
\caption{The conductance as a function of $d_B$ for $N=3$ 
barriers with $E=2$ and $d_0=10$ (solid line). The dashed line is the curve 
given by the classical limit Eq. (\ref{Gs}).}
\label{cond1_db}
\end{figure}

More remarkably we find that, rarefying the magnetic field by adding more 
barriers without changing the total flux $\Phi$, the transmission probability 
approaches the classical limit, where it is zero or one,
depending on whether the incidence angle exceeds or not the 
angular threshold, see Fig. \ref{tran1_N}. Correspondingly, the conductance 
as a function of $N$ approaches the classical limit, Eq. (\ref{Gs}), 
see Fig. \ref{cond1_N}. 
As expected, the same limit is also approached upon increasing the energy, 
especially for large $N$. This is clearly illustrated in Fig. \ref{tran1_E} 
for the transmission, and in Fig. \ref{cond1_E} for the conductance: one sees 
that already for $6$ barriers the classical limit provides a very good 
approximation. 
The classical limit is instead hardly achieved changing $d_B$, 
except when $d_B$ is close to the extreme values $0$ and $E/N$, 
see Figs. \ref{tran1_db} and \ref{cond1_db}. 

In conclusion the main result of this Section is that, 
at fixed flux $\Phi$, the larger is the number of 
barriers, the more transparent is the magnetic structure.
This is a purely quantum mechanical effect, peculiar 
to magnetic barriers. 

Qualitatively this behavior can be explained as follows. 
To calculate the probability for a relativistic particle 
to go through a very thin single barrier, we can 
simulate the profile of $A$ with a step function with 
height $\Phi/N$ in order to have the same flux of a magnetic 
barrier with width $d_B=\Phi/N$. 
In this case the transfer matrix is simply
\begin{equation}
{\hat T}={\cal W}_0(0^-)^{-1}{\cal W}_0(0^+) ,
\end{equation}
where ${\cal W}_0(0^-)$ is given by Eq. (\ref{W_0}) 
with $A=0$ and $k_x=k_x^i$, while in ${\cal W}_0(0^+)$ 
we have $A=\Phi/N$ and consequently $k_x=k_x^f$. 
From Eq. (\ref{transm}) we get the following transmission 
probability for a single barrier 
\begin{equation}
\label{Tau1}
{\cal T}_1(\phi)\simeq \frac{4\cos\phi\,
\cos\phi_f\,\theta(1-|\sin \phi +\Phi/NE|)}{(\cos\phi+
\cos\phi_f)^2+[\Phi/(NE)]^2},
\end{equation}
where $\phi_f$ is defined by $E\sin\phi_f=E\sin\phi+\Phi/N$. 
For $N$ barriers we can roughly estimate the probability 
for the particle to cross the magnetic structure to be  
\begin{equation}
\label{P(N)}
{\cal T}_N(\phi)\sim {\cal T}_1(\phi)^N\,\theta(1-|\sin \phi +\Phi/E|),
\end{equation}
where we have put by hand the global constraint 
of momentum conservation. 
Using the expression above we can then calculate the 
conductance applying Eq. (\ref{cond}). To simplify the 
calculation, in order to have a qualitative description 
of the conductance behavior, we further approximate 
$\cos\phi_f\simeq \cos\phi$ in Eq. (\ref{Tau1}), valid 
for small $\Phi/N$, and replace $\cos\phi^2$ with its 
average $1/2$, getting at the end an approximated 
expression for $G$ which reads 
\begin{equation}
\label{G_approx}
G \simeq G_s \left(\frac{2N^2E^2}{2N^2E^2+\Phi^2}\right)^N,
\end{equation}
being $\Phi$ the total flux for $N$ barriers. 
In Fig. \ref{cond1_N} we compare the exact calculation with the 
approximate one given by Eq. (\ref{G_approx}), and find a 
surprisingly good agreement.
The small discrepancy has various possible sources:
the angular dependence of the transmission, which is averaged 
out in the approximate calculation;
the finite width of the barriers;
the finite separations among the barriers, 
which may let the particles bounce back and forth, 
in this way reducing the transmission. 
This latter effect, therefore, suppresses a 
bit the conductance predicted by Eq. (\ref{G_approx}).

\section{Alternating magnetic barriers and wells}
\label{sec4}

Next, we consider the magnetic profile in Eq. (\ref{gauge2}), 
illustrated in Fig. \ref{profile2}.  
In order to construct the transfer matrix 
in this case we need ${\cal W}_B$ and the
partial transfer matrix for the regions with 
$B_z=-B$, which is given by  
\begin{equation}
{\cal W}_{-B}(x) =
\left(
\begin{array}{ll}
D_{p+1}(-q) & D_{p+1}(q)\\
\frac{-i\sqrt{2}}{E}(p+1)D_{p}(-q) &\frac{i\sqrt{2}}{E}(p+1)D_{p}(q)
\end{array}
\right).
\end{equation}
After some algebra, we then get 
\bea
\hat{T}&=&{\cal W}_{0}^{-1}(x_0) \, {\cal W}_{-B}(x_0) \, 
\hat T_0 \, \hat T_1 \cdots \nonumber \\
&&\quad \cdots \hat T_{N-2}\, \hat T_{N-1} \, 
{\cal W}_{-B}^{-1}(x_{N})\, {\cal W}_0(x_{N}),
\eea
where
\be
\hat T_n = 
{\cal W}_{-B}^{-1}(x_{n})\, 
{\cal W}_{B}(x_{n})\,
{\cal W}_{B}^{-1}(x_{n}+d_B)
\, {\cal W}_{-B}(x_{n}+d_B),
\label{tncase2}
\ee
is the transfer matrix\cite{transfer} 
across the $(n+1)^{\text{th}}$ magnetic barrier, and $x_n=n(d_B+d_{-B})$.
Note that Eq. (\ref{tncase2}) differs from Eq. (\ref{tncase1}),
since now on the right and on the left of a magnetic barrier
there is a magnetic well rather than a nonmagnetic region.

\begin{figure}[h]
\includegraphics[width=0.44\textwidth]{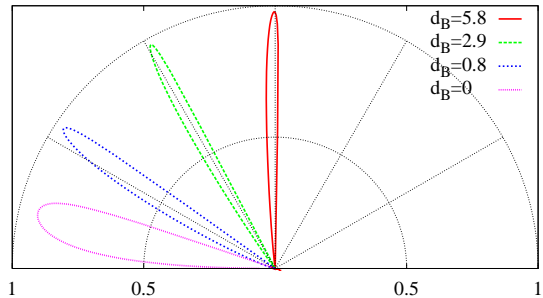}
\caption{(Color online) Angular dependence of the transmission probability 
at $E=6$ through a structure consisting of one magnetic barrier 
of width $d_B$ and one magnetic well of width $d_{-B}$. 
We fix $d_{-B}=11.7\approx 2E$ and vary $d_B$ from $0$ to $5.8\approx E$.}
 \label{tran2_pol}
\end{figure}

\begin{figure}[h]
\includegraphics[width=0.44\textwidth]{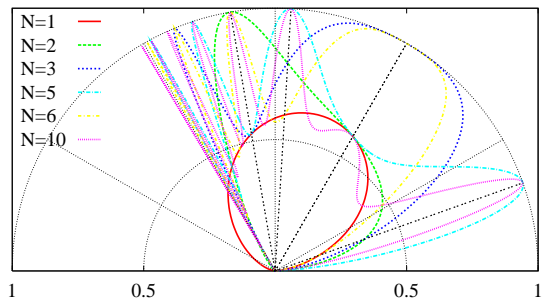}
\caption{(Color online) Angular dependence of the transmission
probability at $E=1$ and $d_B=d_{-B}=1$ 
for several values of $N$. The black dashed lines correspond to
the angles for recursive transmission of multiplicity $2$ (at $\phi\approx \pi/18$), $3$ (at $\phi=-\pi/6$) and $5$ (at $\phi\approx -\pi/54$ and $\phi\approx -7\pi/18$).}
\label{tran2_N}
\end{figure}

\begin{figure}[h]
\includegraphics[width=0.44\textwidth]{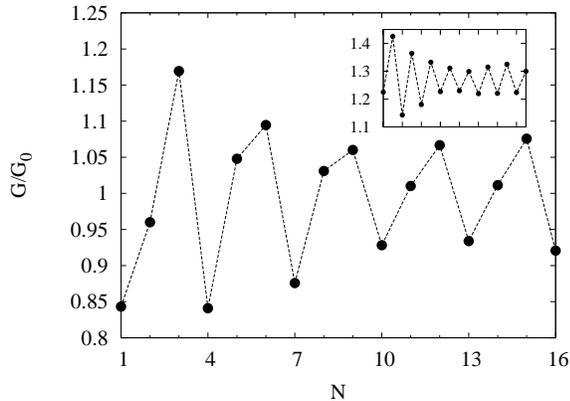}
\caption{The conductance $G/G_0$ at $E=1$ varying $N$ with $d_B=d_{-B}=1$. 
In the inset $d_B=d_{-B}=0.8$ in the same range of $N$.}
\label{cond2_N}
\end{figure}

\begin{figure}[h]
\includegraphics[width=0.44\textwidth]{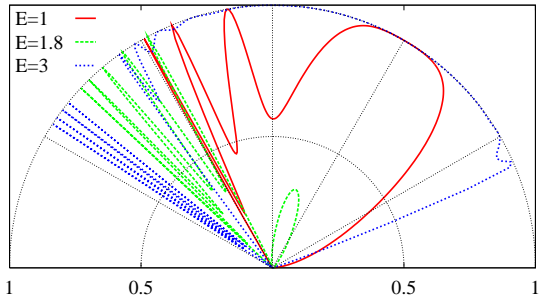}
\caption{(Color online) Angular dependence of the transmission probability 
for several values of the energy $E$, 
for $d_B=d_{-B}=1$ and $N=6$.}
\label{tran2_E}
\end{figure}

\begin{figure}[h]
\includegraphics[width=0.44\textwidth]{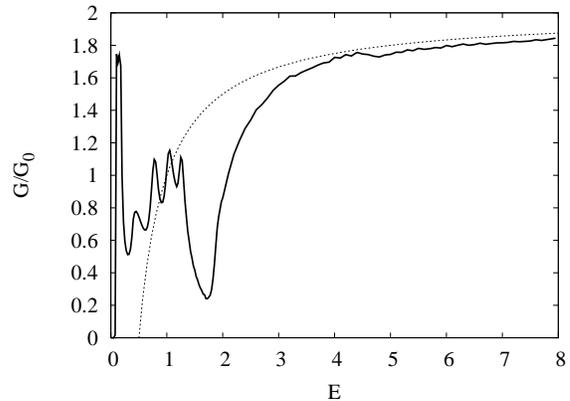}
\caption{The conductance $G/G_0$ as a function of the energy $E$
 for $N=6$ with $d_B=d_{-B}= 1$ (solid line). The dashed line is the classical limit, i.e. $(2-d_B/E)\,\theta(2E-d_B)$.}
\label{cond2_E}
\end{figure}

\begin{figure}[h]
\includegraphics[width=0.44\textwidth]{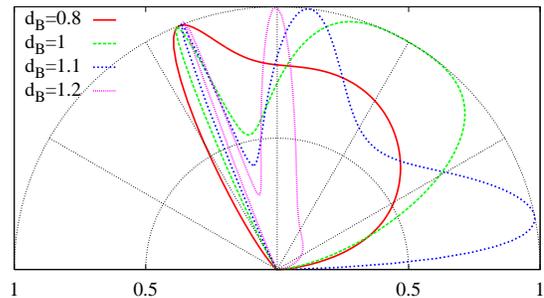}
\caption{(Color online) Angular dependence of the transmission probability
for several values of $d_B=d_{-B}$ with $N=3$ and $E=1$}
\label{tran2_dB}
\end{figure}

\begin{figure}[h]
\includegraphics[width=0.44\textwidth]{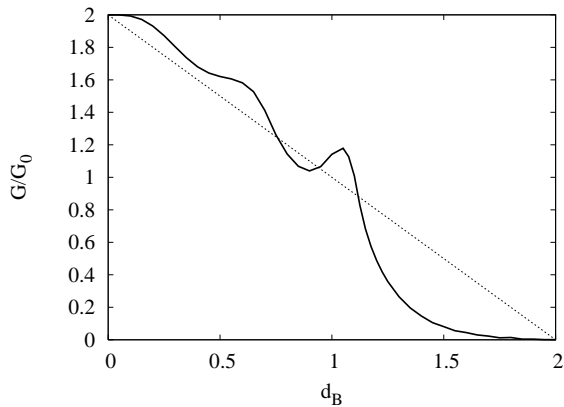}
\caption{The conductance $G/G_0$ as a function of $d_B=d_{-B}$
 for $N=3$ with $E=1$. The dashed line is the classical limit.}
\label{cond2_db}
\end{figure}

As in the previous section, the numerical evaluation of $\hat T$
is straightforward, and the results for the transmission probability 
and for the conductance are illustrated in Figs. \ref{tran2_pol} 
to \ref{cond2_db}.

Fig. \ref{tran2_pol} shows the angular dependence of ${\cal T}$
for a single block consisting of a barrier followed by a well of 
different width. The plot emphasizes the very strong wave-vector 
dependence of the transmission, and shows that by tuning $d_B$ 
and $d_{-B}$ one can achieve very narrow transmitted beams. 
This suggests an interesting application of this structure as a 
magnetic filter, where only quasiparticles incident with an angle 
within a very small range are transmitted.

More explicitly, in order to select a narrow beam at an angle 
$0<\phi< \frac{\pi}{2}$ (at fixed energy),
one can choose the widths such that 
\bea
\label{dB_good}
\frac{d_B}{E} &=& (1-\sin\phi)-\varepsilon_1,\\
\frac{d_{-B}}{E} &=&  2 -\varepsilon_2,
\eea  
with $\varepsilon_{1,2}\ll 1$. 
With this choice, $\phi$ is very close to the angular threshold 
for the first barrier, and the well is close to be totally reflecting. 
The combination of these two effects leads to the narrow beam.
For $-\frac{\pi}{2} < \phi <0$ it is enough to flip the magnetic field. 
These relations hold better at small values of $|\phi|$.

In the rest of this section, we focus on structures with  
$d_B=d_{-B}$, i.e. when the total magnetic flux through 
the structure vanishes.

For several blocks of barriers and wells we observe 
an interesting {\it recursive effect} shown in Fig. \ref{tran2_N}. 
There are angles at which, upon increasing $N$, 
the transmission takes at most $N_m$ values, where $N_m$ is 
the smallest number of blocks for which the transmission is 
perfect (i.e. ${\cal T}=1$). Some of these angles are emphasized 
in Fig. \ref{tran2_N} by dashed lines. At those angles, for instance, 
magnetic structures with $N$ equal to integer multiples of $2$, $3$ 
and $5$ exhibit perfect transmission, even for small energy.

These effect can be understood as follows.
Suppose that, at a given angle, the transmission 
probability through $N_m$ cells has a resonance and 
reach the value 1. Then, for any sequence consisting 
of a number of cells equal to an integer multiple of $N_m$, 
the transmission is 1 again. (It is crucial here that, since the
magnetic flux through each block is zero, the emergence angle always 
coincides with the incidence angle.)
At such an angle, then, $\cal T$ can only take, upon changing $N$,
at most $N_m$ values. 
Notice that perfect transmission occurs also for low energy 
of the incident particle, and the angular spreading of perfect transmission 
is also reduced by adding multiple blocks. 

This effect in the transmission also reflects
in the conductance. Fig. \ref{cond2_N} shows that, for a particular 
set of parameters, $G$ oscillates as a function of $N$ with period $3$.
However,  for a different value of $d_B$, shown in the inset, the period is $2$.
This unexpectedly strong dependence of the conductance 
on adding or removing blocks of barriers and wells could be 
exploited to design a magnetic switch for charge carriers 
in graphene.
Moreover, we observe that the angular dependence of the
transmission is abruptly modified also by changing the energy $E$ 
of the incident particles, see Fig. \ref{tran2_E}, or the width of the 
barriers $d_B$, Fig. \ref{tran2_dB}, where we observe pronounced
resonance effects. As a consequence the conductance exhibits a 
modulated profile as a function of both the energy and the barrier's width, 
as illustrated respectively in Fig. \ref{cond2_E} and \ref{cond2_db}.

\section{Periodic magnetic superlattice}
\label{sec5}

\begin{figure}[h!]
\includegraphics[width=0.32\textwidth]{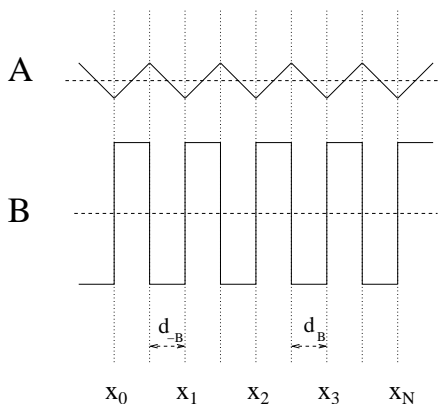}
\caption{Periodic superlattice of alternating magnetic barriers and wells 
with $d_B=d_{-B}$.}
\label{superlattice}
\end{figure}

\begin{figure}[h!]
\includegraphics[width=0.44\textwidth]{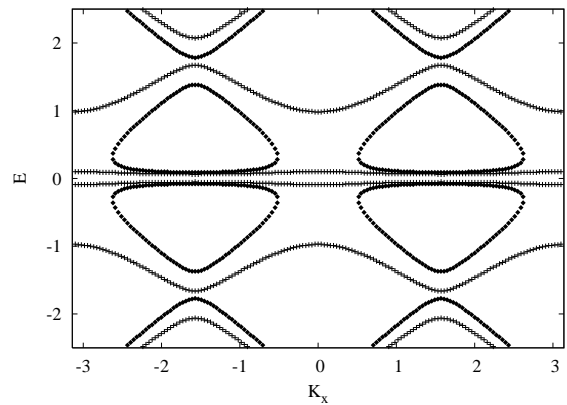}
\caption{The spectrum for the periodic superlattice in Fig. \ref{superlattice}
with $d_B=d_{-B}=1$ at $k_y=0$ (full points) and $k_y=1$ (cross poins). 
$K_x$ is the Bloch momentum.}
\label{spectrum1}
\end{figure}

\begin{figure}[h!]
\includegraphics[width=0.44\textwidth]{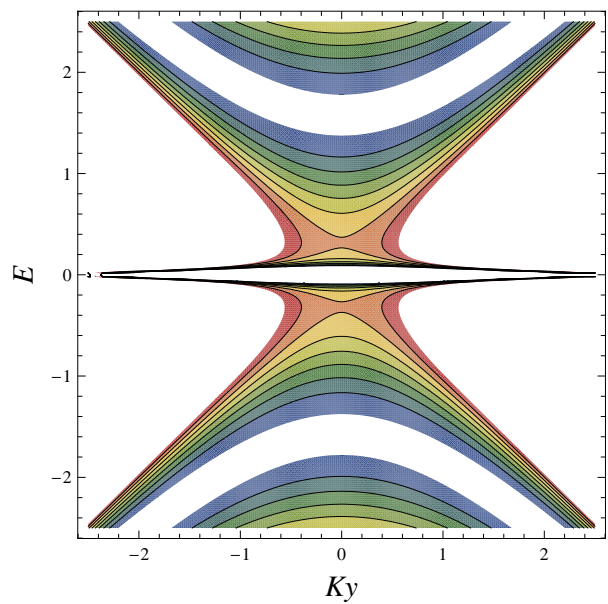}
\caption{(Color online)
The allowed spectrum, 
$|{\text{Tr}}\,\Omega |\le 2$, 
varying $E$ and $k_y$, at $d_B=d_{-B}=1$. 
The contour lines correspond to the values of ${\text{Tr}}\,\Omega$ in 
the interval $[-2,2]$ at steps of $0.5$, 
increasing from blue (inner dark gray) to red (outer dark gray).}
\label{spectrum2}
\end{figure}

We now focus on the case of a periodic magnetic superlattice.
We observe that, if $d_B=d_{-B}$, the profile (\ref{gauge2}) 
can be extended to a periodic profile, illustrated in Fig. \ref{superlattice}, 
where the elementary unit is given by the block formed by a barrier and a well.
Imposing periodic boundary conditions on the wavefunction 
after a length $L=x_N=N \ell$ (where $\ell=2d_B$),
i.e. $\psi(x_0)=\psi(x_N)$, 
and defining the matrix
\be
\Omega={\cal W}_B^{-1}(0){\cal W}_{-B}(0)
{\cal W}_{-B}^{-1}(d_{B}){\cal W}_B(d_B),
\ee
standard calculations\cite{mckellar} lead to the 
quantization condition for the energy:
\be
\label{sc}
2\cos(K_x \ell)={\text{Tr}}\, \Omega .
\ee
At fixed $k_y$, Eq. (\ref{sc}) gives the energy as 
function of the Bloch momentum $K_x=\frac{2\pi n}{L}$. 
Notice that $K_x$ is related to the periodicity of the structure, and 
parametrizes the spectrum. It should not be confused with the 
$x$-component of the momentum $k_x$ used in the previous sections.
Fig. \ref{spectrum1} illustrates the first two bands as a function 
of $K_x$ for two values of $k_y$.
Fig. \ref{spectrum2} shows the contour plot for ${\text{Tr}}\,\Omega$ as 
function of $E$ and $k_y$. We find two interesting main features.
First, around zero-energy there is a gap, whose width decreases for 
larger values of $|k_y|$. This is in agreement with the fact that for a
magnetic profile with zero total flux there exist no zero-energy states.

Second, for some values of $k_y$, the group velocity 
$v_y=\frac{\partial E}{\partial k_y}$ diverges
(see Fig. \ref{spectrum2} close to $|k_y| > |E|\approx 0.3$ ). 
The property of {\em superluminal} velocity has already been 
observed for massless Klein-Gordon bosons in a 
periodic scalar potential\cite{barbier}. However, to our knowledge, this is 
the first example in which such property is observed for massless 
Dirac-Weyl fermions in a periodic vector potential.

\section{Summary and conclusions}
\label{sec6}

In this paper we have studied the transmission 
of charge carriers in graphene through complex 
magnetic structures consisting of several magnetic barriers 
and wells, and the related transport properties.

We focussed on two different types of magnetic profiles.
In the case of a sequence of magnetic barriers, we have 
found that the transparency of the structure is enhanced
when the same total magnetic flux is distributed over 
an increasing number $N$ of barriers. The transmission 
probability and the conductance then approach the classical limit 
for large $N$, see in particular Figs. \ref{tran1_N} and \ref{cond1_N}. 

The behavior of alternating barriers and wells turns out to be 
even more interesting. We have shown that a single unit 
consisting of a barrier and a well of suitable widths can be used as
a very efficient wave-vector filter for Dirac-Weyl quasiparticles, 
see Fig. \ref{tran2_pol}. 
With several blocks we have observed strong resonant effects,
such that at given angles one gets narrow 
beams perfectly transmitted even for low energy of the
incident quasiparticles, see Fig. \ref{tran2_N}. As a result, 
the conductance is drastically modified by adding or removing 
blocks, see Fig. \ref{cond1_N}. This suggests possible applications
as magnetic switches for charge carriers in graphene.

We hope that our paper will further stimulate
experimental work on the rich physics of 
magnetic structures in graphene. 

\acknowledgments

We thank R. Egger and W. H\"ausler for several
valuable discussions.  
This work was supported by the MIUR project ``Quantum noise in mesoscopic systems'', 
by the SFB Transregio 12 of the DFG and by the ESF network INSTANS.


\begin{thebibliography}{99}
\bibitem{footnote1} In all cases discussed in this paper, the 
transmission probability is the same at the two valleys.
\bibitem{transfer} Note that, according to the usual definition, 
our $\hat T$ is actually the inverse of the transfer matrix.
\end{thebibliography}
\end{document}